\documentclass[a4paper,12pt]{article}

\usepackage{multicol}
\begin{document}
\title{\vskip -70pt
\begin{flushright}
{\normalsize \ DAMTP-97-97}\\
\end{flushright}
\vskip 20pt
{\bf Vortices and Flat Connections }
}
\vspace{1cm}
\author{{S. M. Nasir}\thanks{e-mail address:
\tt {S.M.Nasir@damtp.cam.ac.uk}}\\
{\sl Department of Applied Mathematics and Theoretical Physics}\\
{\sl University of Cambridge} \\
{\sl Silver Street, Cambridge CB3 9EW, England}}
\date{July, 1998}
\maketitle
\vspace{-.5cm}
\begin{abstract}
\noindent
At Bradlow's limit,
the moduli space of Bogomol'nyi vortices on a compact Riemann
surface of genus $g$ is determined.
The
K\"{a}hler form, and the volume of the moduli space is then computed.
These results are compared with the corresponding results
previously obtained for a general vortex moduli space.

\end{abstract}
\vspace{.5cm}
\begin{multicols}{2}


\noindent
{\bf{1.}} The Abelian Higgs model in (2+1) dimensions
is an interesting arena to study
vortices. The coupling constant of the model determines the nature of
interactions among vortices.
At the critical
coupling, the model
admits static and finite energy Bogomol'nyi vortex
solutions \cite{Bog}.
Stability of these solutions is ensured by topology. We will consider
vortices in a
space-time of the form ${\bf{R}}\times M,$ where $M$ is a compact two
dimensional manifold. 
The metric of the space-time is taken to be
$ds^{2}=dx_{0}^{2}-\Omega (x_{1},x_{2})(dx_{1}^{2}+dx_{2}^{2})$, where
$x_{1}$ and $x_{2}$ denote local coordinates on $M$.
Let $A_{\mu},\;
(\mu=0,1,2)$ be a $U(1)$ gauge potential and $\phi$ be a complex scalar
field. 
Working in the gauge $A_{0}=0$, the Lagrangian of the model at the
critical coupling is $L=T-V$, where
\begin{equation}\label{KE}
T=\frac{1}{2}\int_{M}
d^{2}x \,
(\dot{A}_{i}\dot{A}_{i}+\Omega \dot{\phi}\dot{{\bar{\phi}}}),\; \; (i=1,2)
\end{equation}
\begin{equation}\label{PE}
\begin{array}{lll}
V & =\frac{1}{2}\int_{M} d^{2}x \Omega  & \left[  \frac{1}{2}F_{ij}F^{ij}
+D_{i}\phi\overline{D^{i}\phi} \right. \\ 
&  & \left. + \frac{1}{4}(|\phi|^{2}-1)^{2}\right]
\end{array}
\end{equation}
are respectively, the kinetic  and the potential energies. Here,
$D_{i}=\partial_{i}-iA_{i},$ and
$F_{12}=\partial_{1}A_{2}-\partial_{2}A_{1}$, is the magnetic field.
The following first order Bogomol'nyi vortex
equations are obtained by minimizing the
potential energy 
\begin{equation}
(D_{1}+ iD_{2})\phi=0
\end{equation}
\begin{equation}
F_{12} + \frac{\Omega}{2}(|\phi|^{2}-1)=0.
\end{equation} 
The above equations admit static multi-vortex solutions. 
The solutions are parametrized by a $2N$
dimensional moduli space, $M_{N}$,
where $N$ is the number of zeros of the Higgs
field counted with multiplicity \cite{Tau, Jaf}.
$N\,-$ called the vortex number $-$
is related to the
total magnetic flux by $N={\displaystyle{\frac{1}{2\pi}\int_{M}d^{2}x
F_{12}}}$. The potential
energy of a configuration of static $N$-vortices is $\pi N$.
Topologically, $M_{N}$ is just the symmetrized $N$-th power
of $M$. The moduli space has a natural Riemannian metric
induced from the kinetic energy expression (1). This is given by
\begin{equation}
ds^{2}=\frac{1}{\pi}\int_{M} d^{2}x (\delta A_{i}\delta A_{i}+\Omega
\delta \phi \delta {\bar{\phi}}), \, \, (i=1,2).
\end{equation}
In obtaining (5) from (1) we have multiplied (1) by $2/\pi$ to agree
with the conventions of \cite{Man2}.
It is to
be noted that there is a Gauss' law constraint arising from the
equation of motion of $A_{0}$. This constraint ensures that $M_{N}$
consists of the solutions of eqns.(3, 4) modulo 
gauge transformations connected
to the identity. Hereafter, by `gauge transformations' we will mean
those gauge transformations connected to the identity.
Through Manton's work \cite{Man1} it is known that at low
energy $-$ when most of the degrees of freedom remain unexcited $-$ the
moduli space can be used to describe interesting physical phenomena
associated with vortices, such as
scattering \cite{Sam}, thermodynamics \cite{Man2, Sha, Nas}, and the phase
transition of vortices at 
near-critical coupling \cite{Sha1}, etc. In the moduli space approximation
the vortex dynamics can be thought
of as geodesic motion on the moduli space.

For a compact surface, in order for vortex solutions to exist one has
to satisfy Bradlow's bound \cite{Bra}.
The bound is $4\pi N\leq A$, where $A$ is the area of
$M$. This can be
obtained by integrating (4) over $M$, and noticing that the integral
of $|\phi|^{2}$ over $M$ is positive.
The bound means that for a given area
there is a limit on the number of vortices one can have on $M$. 
At Bradlow's limit, $A=4\pi N$,
the Higgs field $\phi$ must vanish everywhere on $M$.
Then, the Bogomol'nyi equations reduce to the following single equation
\begin{equation}
F_{12}=\frac{\Omega}{2}.
\end{equation}
This is an equation for a constant magnetic field on $M$. The energy
of the configuration is still $\pi N$.

It might be thought that (6) has a unique solution up to
gauge equivalence, which in turn will mean that the moduli space is just a
point. However, if $M$ has non-contractible loops (one-cycles),
i.e. if the first
homotopy group of $M$
is non-trivial, then
the moduli space of solutions of (6) is non-trivial as well. 
A solution of (6) up to gauge equivalence is given, in addition to
a gauge potential
that solves (6), by specifying holonomies
around a basis of one-cycles. 
The holonomies around any two homologous loops
are generally different as the magnetic field is non-zero on $M$, but
the difference can be completely determined by using Stokes' theorem. Now,
linearizing (6) around a particular solution one can see that the
perturbed gauge potentials satisfy the equations for flat $U(1)$ connections
for which the magnetic field is zero. Flat connections are associated
with large gauge transformations.
Flat connections up to gauge equivalence are also given by
specifying their holonomies around a basis of one-cycles. 
Hence, when $\phi=0$ the moduli space of Bogomol'nyi
vortices is no longer $M_{N}$. Locally, the tangent space of the
vortex moduli space at Bradlow's limit
can be identified with the tangent space 
of the space of flat $U(1)$ connections on $M$. With an abuse of
notation, in future 
the moduli space of solutions of (6) and the space of flat $U(1)$
connections will be denoted by
$M_{f}$.  
It was demonstrated long ago by Aharonov and Bohm \cite{AB} that flat
connections play quite a non-trivial role in quantum physics.
It should be noted that flat connections do not contribute to the
moduli of the Bogomol'nyi equations when the Higgs field is
non-vanishing.

In
passing we would like to point out that when $\phi=0$, one may consider
$F_{12}=0 $ as solutions of the static Abelian Higgs model. The
energy is then $A/8$ with no restriction on the value of $A$. 
However, these solutions cannot be obtained from Bogomol'nyi equations.   
Henceforth, we will only consider vortices at Bradlow's limit, and
$M$ is taken to be of genus
$g\geq 1$.

\noindent
{\bf{2.}} Let $A=A_{1}\, dx_{1}+A_{2}\, dx_{2}$ solve (6). Then,
$A+\delta A$, where $\delta A=
a_{1}dx_{1}+ a_{2}dx_{2}$ is a flat
connection, is also a solution of (6). The equation satisfied by
$\delta A$ is
$d\delta A=0$. 
The metric on $M_{f}$ is
obtained by putting $\phi =0$ in (5).
This is
\begin{equation}
ds^{2}=\frac{1}{\pi}\int_{M} d^{2}x [( a_{1})^{2}+(a_{2})^{2}].
\end{equation}
Notice that the metric of $M$ does not appear in the above expression.
This means that the metric information of $M$ is not carried
over to $M_{f}$.
One can see that $M_{f}$ inherits a complex structure from $M$.
The map ${\mathcal{I}}$ given by ${\mathcal{I}}:
a_{j}\rightarrow
-\epsilon_{jk}a_{k}, $, where $\epsilon_{jk}$ is the
antisymmetric tensor with $\epsilon_{12}=1$,
leaves invariant (7) and the equation for a flat connection. Moreover,
${\mathcal{I}}^{2}=-{\bf{1}}_{2}$. Hence, ${\mathcal{I}}$
defines an almost complex structure on $M_{f}$.
This almost complex structure can be used to
define the following (1,1) form on $M_{f}$ 
\begin{equation}
\omega (\delta A, \delta B)=\frac{1}{\pi}\int_{M}(\delta A_{z}\wedge
{\overline{\delta B_{z}}}-\delta B_{z}\wedge
{\overline{\delta A_{z}}}).
\end{equation}
Here, we have used complex coordinates $(z=x_{1}+ix_{2})$, and $\delta
A =(\delta A_{z}+c.c)$. 
Clearly, $\omega$ is
manifestly real. Using $d\delta A=d\delta B=0$, it can also be shown
that $\omega$ is
a closed form.
Thus, ${\mathcal{I}}$ is a complex structure and $\omega $ defines
a K\"{a}hler form. Among many uses
of this K\"{a}hler form one can, for example, compute the volume of 
$M_{f}$.
Recently, we obtained an expression for the K\"{a}hler form on $M_{N}$
and also, we computed the volume of $M_{N}$
\cite{Nas}. We will see that when $A=4\pi
N$, the K\"{a}hler form on $M_{f}$
gets mapped to the K\"{a}hler form on $M_{N}$.

Although a known fact, in order to setup the stage for the main part
of this paper
we intend to show that the space of flat $U(1)$
connections on $M$
modulo gauge transformations, i.e. $M_{f}$,
is parametrized by the Picard variety, ${\tilde{J}}$, of
$M$ \cite{Gun}. ${\tilde{J}}$ is dual to the Jacobian $J$ of
$M$. 
$J$ is a $2g$-dimensional
real torus. 
Let $\nu_{i},\; (i=1,\cdots ,2g)$ be a basis of $2g$
one-cycles of $M$. 
Let $w_{\rho},\; (\rho =1,\cdots ,g)$ be a basis of $g$
holomorphic one-forms on $M$. 
Define the period matrix $W=(w_{\rho i}),$ where $w_{\rho
i}=\oint_{\nu_{i}} w_{\rho},\; (\rho =1,\cdots ,g,\;
i=1,\cdots ,2g)$. The columns $W_{i}, \; (i=1,\cdots ,2g)$
of $W$ can be thought of as spanning a $2g$-dimensional
lattice in ${\bf{C}}^{g}$. $J$ is defined as
the torus ${\bf{C}}^{g}/W$. Riemann's bilinear
relations can be used to normalize the period matrix $W$ \cite{GH}.
One can choose $W$ such that
$(W^{t}, \, {\bar{W}}^{t})$ is a symmetric matrix
with $Im(W)>0$. Further, the elements of $W$ can be restricted to
satisfy $w_{\rho i}=\delta _{\rho i}$ for $\rho =1,\cdots
,g$ and $i=1,\cdots ,g$, the remaining elements being arbitrary with
positive imaginary parts \cite{GH}. 
Let us choose a basis of $2g$ one-cocycles $\alpha_{i}$
such that $\oint_{\nu_{i}} \alpha_{j}=\delta
_{ij},\; (i,j=1,\cdots ,2g)$. These one-cocycles are the $2g$
generators of the cohomology group $H^{1}(M,{\bf{Z}})$. 
A canonical basis of $\nu_{i}$ can be chosen such that the cocycles
also satisfy ${\displaystyle{\int_{M} \alpha_{i}
\alpha_{j}=\delta_{i,j+g}}},$ where in the
integration wedge product is implied. The $g$ holomorphic
one-forms can be expressed in terms of $\alpha_{i}$ as $w_{\rho}=
{\displaystyle{\sum_{i=1}^{2g}}}
w_{\rho i}\alpha_{i}$ for $(\rho =1,\cdots ,g)$.
Reciprocally, $\alpha_{i}$ can be expressed in terms of $w_{\rho}$ as
$\alpha_{i}={\displaystyle{\sum_{\rho=1}^{g}(\gamma_{i\rho}w_{\rho}+c.c)}}$,
for
$(i=1,\cdots ,2g)$ where the matrix $\Gamma =(\gamma_{\rho i})$ is
required to
satisfy
\begin{equation}
\Gamma^{t} W+{\overline{\Gamma^{t}}}\, {\overline{W}}={\bf{1}}_{2g}.
\end{equation}
We note that $\tilde{J}$ is defined as the torus ${\bf{C}}^{g}/\Gamma$.
Hence, $\tilde{J}$ is dual to $J$.

Let $c_{\rho},\; (\rho =1,\cdots
,g)$ denote complex coordinates on $M_{f}$ (that $M_{f}$ is
$2g$-dimensional will be evident below).
Then a real flat connection 
$A_{f}$ (modulo gauge transformations) can be expressed as 
\begin{equation}
A_{f}=2\pi \sum_{\rho=1}^{g}(c_{\rho}
w_{\rho}+\; c.c).
\end{equation} 
Imposing Gauss' law on the flat connections one can see that the
above is the most general expansion for a flat connection. 
The holonomy, $h_{j},$ of $A_{f}$
around a one-cycle $\nu_{j}$ is given by
\begin{equation}
h_{j}=\exp
(i\oint_{\nu_{j}}A_{f})=\exp(2\pi i\sum_{\rho=1}^{g}(c_{\rho}w_{\rho
j}+\; c.c)).
\end{equation}
As noted earlier, for a complete specification of
$A_{f}$ one needs to specify all of
the holonomies $h_{j}$.
Eqn.(11) implies that $h_{j}$
is periodic with the period matrix being $\Lambda=
(\lambda_{\rho i}),
\;(\rho=1,\cdots ,g,\; i=1,\cdots ,2g)$, say. 
Then the $2g$ columns of $\Lambda$ span a $2g$-dimensional
lattice in ${\bf{C}}^{g}$. 
Thus, $M_{f}$ is parametrized by a
$2g$-dimensional real torus ${\bf{C}}^{g}/\Lambda$.
Further, 
the following relation for $\Lambda$ is implied by (11) 
\begin{equation}
\Lambda^{t} W+{\overline{\Lambda^{t}}}\, {\overline{W}}={\bf{ 1}}_{2g}.
\end{equation}
Comparing the above equation with (9) one
gets $\Lambda=\Gamma$. This identifies $M_{f}$ with
$\tilde{J}$.

There are $g$ independent holomorphic one-forms on $M_{f}$.
Using $c_{\rho}$ as coordinates of $M_{f}$, these are
given by
$dc_{\rho},\;(\rho=1,\cdots ,g).$ A basis of one-cycles in $M_{f}$ is
given by the $2g$ lines $t\Lambda_{i}$, $0\leq t \leq 1$, with $\Lambda_{i}$
identified with 0 to produce a closed loop.
Then from the discussion preceeding (9), one deduces that the $2g$
generators of
$H^{1}(M_{f},{\bf{Z}})$ are given by
$\tilde{\xi}_{i}={\displaystyle{\sum_{\rho=1}^{g}(w_{\rho
i}dc_{\rho}+c.c)}}$. Now from (8), we get the following K\"{a}hler
form on $M_{f} $
\begin{equation}
\tilde{\omega} =4\pi \sum_{\rho ,\rho ',i=1}^{g}(w_{i \rho }
{\bar{w}}_{i+g \rho'
}-{\bar{w}}_{i \rho' }
w_{i+g \rho
}) dc_{\rho}\wedge d{\bar{c}}_{\rho '}.
\end{equation}
In terms of ${\tilde{\xi}}_{i}$, $\tilde{\omega}$ can be written as
\begin{equation}
\tilde{\omega} =4\pi \sum_{i=1}^{g} {\tilde{\xi}}_{i}{\tilde{\xi}}_{i+g}.
\end{equation}
In obtaining the above from (13) we have used the Riemann bilinear relations.
The volume of $M_{f}$ is then
\begin{equation}
{\rm{Vol}}_{f}=\frac{1}{g!}\int_{M_{f}}\tilde{\omega}^{g}=(4\pi )^{g}
\end{equation}
where use has been made of the fact that
{\mbox{${\displaystyle{\int_{M_{f}} \prod
_{i=1}^{g}\tilde{\xi}_{i}\tilde{\xi}_{i+g}}}  =1$}}. It is useful to
notice that the volumes of $J$ and $\tilde{J}$ are the same.
The computation of the volume of the space of flat connections on a compact
Riemann
surface is not
new. For $SU(2)$ and $SO(3)$
Yang-Mills theory, Witten \cite{Witten}
computed the volume of the space of flat connections by a remarkable
use of the Verlinde formula \cite{Ver} in conformal field
theory.
 
\noindent
{\bf{3.}} The general formula for the K\"{a}hler form on the vortex
moduli space $M_{N}$, when $\phi$ is non-vanishing, is \cite{Nas}
\begin{equation}\label{100}
\omega =(A-4\pi N)\eta +4\pi \sum_{i=1}^{g}\xi_{i}\xi_{i+g}
\end{equation}
where $\eta$ is an area form on $M_{N}$ normalized to unity and
$\xi_{i},\; (i=1,\cdots ,2g)$ are the $2g$ generators of
$H^{1}(M_{N},{\bf{Z}})$.
When $\phi =0$ is zero, i.e. $A=4\pi N$, the
K\"{a}hler form in (16) reduces to
\begin{equation}
\omega =4\pi \sum_{i=1}^{g}\xi_{i}\xi_{i+g}.
\end{equation}
At this point we should remind the reader
that this K\"{a}hler form is defined on
$M_{N},$ not on $\tilde{J}$ 
whose tangent space coincides with the tangent
space of the moduli space of vortices when $\phi=0$.
However, it can be shown that the K\"{a}hler form $\omega$ on $M_{N}$
is mapped in a one-to-one way to the K\"{a}hler form $\tilde{\omega}$ on
$J$. As $\tilde{J}$ is isomorphic to $J$ it is enough to
establish an isomorphism between 
$H^{1}(M_{N},{\bf{Z}})$ and
$H^{1}(J,{\bf{Z}})$. First, notice that Jacobi's inversion
theorem \cite{GH} implies that there is an isomorphism
between $J$ and $M_{g}$ where $M_{g}$ is the moduli space of $g$
vortices, and $g$ is the genus of $M$. This implies the isomorphism
between $H^{1}(J,{\bf{Z}})$ and
$H^{1}(M_{g},{\bf{Z}})$. Next, using the Lefschetz hyperplane section theorem
\cite{GH}, one sees that there is an isomorphism between
$H^{1}(M_{N},{\bf{Z}})$ and $H^{1}(J,{\bf{Z}})$ for $N\geq g$ and $g>1$. 
The isomorphism for other values of $N$ and $g$ can also be easily established
by arguments used in \cite{Mac}.

It is of interest to see if one can relate ${\rm{Vol}}_{f}$ to the volume
of $M_{N}$ near Bradlow's limit when $\epsilon =A-4\pi N$ is a small
positive quantity. 
For genus $g\geq 1$ and $N\geq 2g-1$, $M_{N}$ has a bundle
structure, where the base is $J$,
and the fibre is $CP_{N-g}$. For
$N\leq g$, $M_{N}$ is analytically homeomorphic to a
$2N$-dimensional submanifold of the Jacobian. Generically, the volume
of $M_{N}$ is not just a product of the volume of the base and the volume
of the fibre.  
The volume of $M_{N}$ as computed in \cite{Nas} is
\begin{equation}
{\rm{Vol}}_{N}=(A-4\pi
N)^{N-g}\sum_{i=0}^{g}\left(\frac{(4\pi)^{i}(A-4\pi N)^{g-i}g!}{(N-i)!
(g-i)!i!}\right) .
\end{equation}
In this formula $N\geq g$. It is easy to write an analogous formula
for $N<g$. Near Bradlow's limit 
the above volume can be written as
\begin{equation}
{\rm{Vol}}_{N}=(4\pi)^{g}\frac{\epsilon^{N-g}}{(N-g)!}
+O(\epsilon^{N-g+1}).
\end{equation}
Neglecting the higher order corrections the above can be written as 
\begin{equation}
{\rm{Vol}}_{N}={\rm{Vol}}_{f}\times \frac{\epsilon^{N-g}}{(N-g)!}
\end{equation}
where the factor $\epsilon^{N-g}/(N-g)!$ can be thought of 
as a contribution coming from the fibre $CP_{N-g}$. Indeed, using
(\ref{100}) and the cohomolgy class of the fibre one can show that the
volume of the fibre is $(A-4\pi N)^{N-g}/(N-g)!$.

In conclusion, we would like to clarify the following
apparent puzzle. In computing the K\"{a}hler form (16),
one needs to extract the non-singular parts of the
expressions like $\partial_{z}\log|\phi|^{2}$
around the zeros of $\phi$. This, however, does not invalidate
the derivation of (16) when $\phi=0$ as one may think.
From the Bogomol'nyi equations one
can always express $\partial_{z}\log|\phi|^{2}$ in terms of the gauge
potentials, which can in principle be used to derive (16) regardless of
whether $\phi$ is zero or not.

\vspace{.3cm}
{\bf{Acknowledgements:}}
\vspace{.25cm}
\noindent
I would like to thank my supervisor Dr N. S. Manton 
for his continuous guidance and very helpful discussions. 
Also thanks
to P. Irwin for critical
comments on this manuscript. This work
was supported by the Overseas Research Council,
the Cambridge Commonwealth Trust and Wolfson college.

\end{multicols}


\begin{thebibliography}{18}

\bibitem{Bog} E. B. Bogomol'nyi, Sov. J. Nucl. Phys. 24 (1976) 449. 

\bibitem{Tau} C. H. Taubes, Comm. Math. Phys. 72 (1980) 277. 


\bibitem{Jaf} A. Jaffe and C. H. Taubes, Vortices and Monopoles
(Birkh\"{a}user, Boston, 1980).

\bibitem{Man2} N. S. Manton, Nucl. Phys. B 400 (1993) 624. 

\bibitem{Man1} N. S. Manton, Phys. Lett. B 110 (1982) 54.

\bibitem{Sam} T. M. Samols, Comm. Math. Phys. {\bf{145}} 149 (1992); Ph.D
Thesis, (Cambridge University 1990) Unpublished.

\bibitem{Sha} P. A. Shah and N. S. Manton, J. Math. Phys. 35 (1994)
1171.

\bibitem{Nas} N. S. Manton and S. M. Nasir, To appear in Comm. Math. Phys.
hep-th/9807017

\bibitem{Sha1} P. A. Shah, Nucl. Phys. B 438 (1995) 589

\bibitem{Bra} S. Bradlow, Comm. Math. Phys. 135 (1990) 1. 


\bibitem{AB} Y. Aharonov and D. Bohm, Phys. Rev. 115 (1959) 485. 


\bibitem{Gun} R. C. Gunning, Lectures on Riemann Surfaces
(Princeton University Press, New Jersey, 1966).

\bibitem{GH} P. Griffiths and J. Harris, Principles of Algebraic
Geometry (John Wiley \& Sons, Inc., New York, 1978).


\bibitem{Witten} E. Witten, Comm. Math. Phys. 141 (1991) 153.

\bibitem{Ver} E. Verlinde, Nucl. Phys. B 300 (1988) 360. 

\bibitem{Mac} I. G. MacDonald, Topology 1 (1962) 319. 

\end{thebibliography}
\end{document}